# Measurement of the nonlinear refractive index of air constituents at mid-infrared wavelengths

S. Zahedpour, J. K. Wahlstrand, and H. M. Milchberg

Institute for Research in Electronics and Applied Physics, University of Maryland, College Park, MD 20742

We measure the nonlinear refractive index coefficients in $N_2$, $O_2$ and Ar from visible through mid-infrared wavelengths ($\lambda$ = 0.4 – 2.4 μm). The wavelengths investigated correspond to transparency windows in the atmosphere. Good agreement is found with theoretical models of $\chi^{(3)}$. Our results are essential for accurately simulating the propagation of ultrashort mid-IR pulses in the atmosphere.

Filamentary propagation of intense ultrashort laser pulses in atmosphere for $\lambda < \sim 1$ μm has been a subject of extensive study [1, 2], finding applications in the generation of terahertz radiation [3], high harmonic generation [4], air lasing [5-7], and remote inscription of optical waveguides into air [8–13]. It has been anticipated that new regimes of laser filamentation are possible at longer wavelengths, such as in the mid-infrared (mid-IR, $\lambda \sim$ 1.5-10 μm), where beam collapse arrest may occur through harmonic walk-off rather than plasma-induced refraction [14]. Mid-IR filamentation is effective for generating coherent keV-level photon beams in high pressure gas-filled capillaries [15] and broad mid-IR supercontinua in high pressure gas volumes [16].

Accurate values for nonlinear coefficients are essential for high fidelity simulations of intense laser propagation [17]—such simulations are indispensable not only for designing experiments and informing applications, but they also motivate the design and parameters of the lasers themselves. We recently showed that in the near-infrared (pump wavelength $\lambda_e$ = 0.8 μm), air propagation simulations depend very sensitively on the values used for the coefficients ($n_2$) describing the instantaneous electronic nonlinear response of air constituents [18]. Best agreement of the simulations with axially resolved measurements occurs for $n_2$ coefficients measured in [19].

In this Article, we present measurements of $n_2$ for the air constituents $N_2$, $O_2$ and Ar at pump wavelengths ranging from 400 nm to 2400 nm. The near through mid-IR wavelengths chosen ($\lambda_e$ = 1250 nm, 1650 nm, 2200 nm, 2400 nm) are within the transparency windows of air [20]. We find that the nonlinear response is quite dispersionless over the range of wavelengths investigated, except near $\lambda_e$=400 nm, consistent with a simple model for the third-order nonlinear susceptibility developed by Bishop [21].

The experimental setup is shown in Fig 1. Pulses from a 1 kHz Ti:Sapphire regenerative amplifier centered at 800 nm are split, with 2.8 mJ pumping an optical parametric amplifier (OPA) [22], which is tunable from 1100 nm to 2600 nm. A chopper reduces the pulse repetition rate to 500 Hz. The remaining portion of the 800 nm pulse is attenuated and weakly focused in a 2 atm xenon gas cell, where filamentation generates a $\lambda$=500–700 nm broadband supercontinuum (SC) transmitted through the pump-rejecting dichroic splitter. A Michelson interferometer splits the SC into two collinear pulses (reference and probe) temporally separated by 2 ps. The dispersive glass in the SC beam path introduces equal positive chirp to ~1.5 ps on the reference and probe. The OPA output pulse and the SC reference/probe are focused in a backfilled chamber, crossing at 2°.

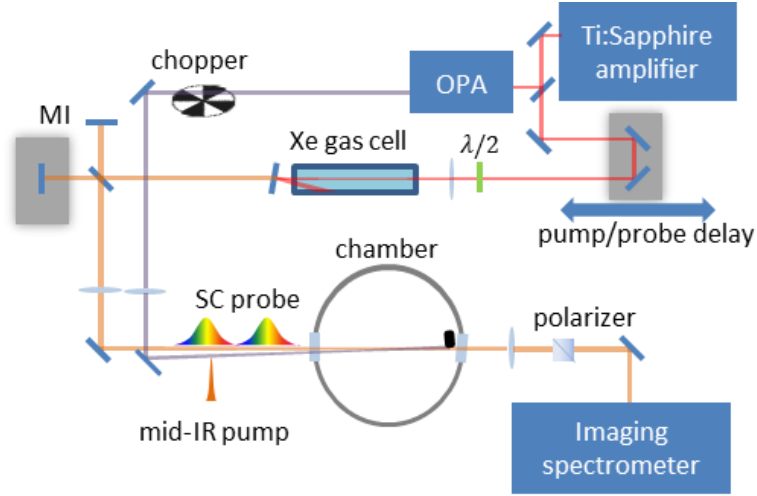

**Fig. 1.** Diagram of the experiment. A Ti:sapphire regenerative amplifier pumps an infrared optical parametric amplifier (OPA) to produce a tunable pump pulse and generates visible supercontinuum (SC) in a Xe gas cell. A Michelson interferometer (MI) creates probe and reference SC pulses. Pump and probe/reference beams are crossed in a chamber filled with $N_2$, $O_2$, or Ar. The spectral phase and amplitude of the probe is measured and used to find the time-domain phase shift induced by the pump pulse.

The crossing angle is small enough to consider all pulses as collinear in the analysis. The reference pulse precedes the pump, which is temporally overlapped by the probe, which is encoded with the pump-induced time-varying nonlinear phase shift. The focal plane of the reference/probe is imaged onto the entrance slit of an imaging spectrometer, inside of which the reference and probe pulses interfere in the spectral domain. Two-dimensional spectral interferograms (1D space and wavelength) are recorded by a CCD camera at the spectrometer's imaging plane. The 2° angular separation between the pump and reference/probe ensures that the pump is stopped at a beam dump before the entrance slit. Fourier analysis of the spectral interferogram, using the measured spectral phase of the probe [23, 24], enables extraction of the time- and 1D space-resolved phase shift $\Delta\varphi(x,t)$ induced by the pump pulse with the time resolution limited to <5 fs by the SC bandwidth. Here, $x$ is a transverse coordinate in the pump focal plane.

The reference/probe polarization is oriented either parallel or perpendicular to the pump polarization by a half waveplate before the xenon SC cell. A Glan-Taylor polarizer ($10^5$:1 extinction ratio) is placed in the reference/probe path to further refine the linear polarization. Pump-probe walk-off is minimal in these experiments: for the probe central wavelength $\lambda_p = 600$ nm and pump wavelength range $\lambda_e = 400$–$2400$ nm, the walkoff for our pump-probe interaction length of ~2 mm is < 0.5 fs, well below the time scale of the most rapid index transients in the experiment, which are of order the pulse width of 40 fs (at $\lambda_e = 800$ nm).

In $N_2$ and $O_2$, the nonlinear response is dominated by the near-instantaneous electronic (Kerr) and the delayed rotational responses [24]. If $\Delta\varphi_\parallel(x,t) = \Delta\varphi_{elec}(x,t) + \Delta\varphi_{rot}(x,t)$ is the SSSI-extracted phase shift for the parallel polarized probe, then nonlinear susceptibility tensor symmetry [25] implies that $\Delta\varphi_\perp(x,t) = \frac{1}{3}\Delta\varphi_{elec}(x,t) - \frac{1}{2}\Delta\varphi_{rot}(x,t)$ for the perpendicular polarized probe. These equations then yield

$$\Delta\varphi_{elec}(x,t) = 3(\Delta\varphi_\parallel(x,t) + 2\Delta\varphi_\perp(x,t))/5$$
$$\Delta\varphi_{rot}(x,t) = 2(\Delta\varphi_\parallel(x,t) - 3\Delta\varphi_\perp(x,t))/5 \quad (1)$$

for the separate electronic and rotational nonlinear responses.

Figures 2(a) and 2(b) show $\Delta\varphi_\parallel(x,t)$ and $\Delta\varphi_\perp(x,t)$ phase shift measurements in nitrogen at $\lambda_e = 1250$ nm pump wavelength. Central lineouts $\Delta\varphi_\parallel(0,t)$ and $\Delta\varphi_\perp(0,t)$ are shown in (c), along with the electronic and

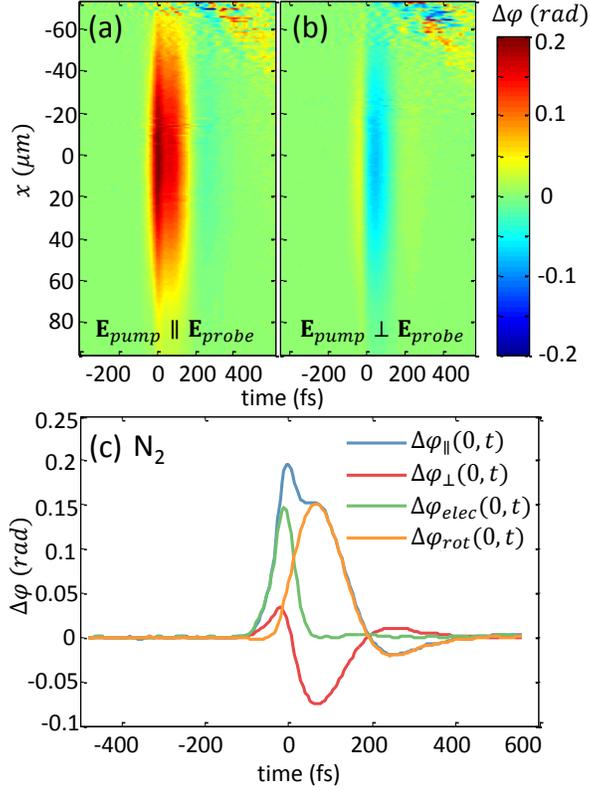

**Fig. 2.** Experimental results in N$_2$ at $\lambda_e$ =1250 nm central pump wavelength. (a) Nonlinear phase shift $\Delta\varphi_\parallel(x,t)$ of the probe for $\mathbf{E}_{pr} \parallel \mathbf{E}_{pump}$ and (b) nonlinear phase shift $\Delta\varphi_\perp(x,t)$ of the probe for $\mathbf{E}_{pr} \perp \mathbf{E}_{pump}$ (c) Temporal lineouts $\Delta\varphi_\parallel(0,t)$ and $\Delta\varphi_\perp(0,t)$ and their decomposition into electronic and rotational responses, as described in the text.

Rotational responses $\Delta\varphi_{elec}(0,t)$ and $\Delta\varphi_{rot}(0,t)$ extracted using Eq. (1). Figure 3 shows the same plot for argon, which lacks a rotational response, verifying that $\Delta\varphi_{elec,\parallel}(0,t) = 3\Delta\varphi_{elec,\perp}(0,t)$ in accord with the symmetry of an instantaneous isotropic nonlinearity. This result verifies the sensitive ability of our technique to separate the electronic and rotational responses in molecular gases.

Our prior absolute determination of the Kerr and rotational nonlinearities at $\lambda$ = 800 nm used auxiliary interferometric measurements of the optical thickness of our thin gas target and measurements of the pump intensity profile [19]. Here, we use an alternative method, employed in other recent pump-probe nonlinearity measurements [26, 27], in which we reference all of our nonlinear phase shift measurements to the rotational responses in nitrogen and oxygen without explicit need for either gas density or pump intensity profiles. We use the fact that, to second order in the pump field, the response, as measured by the nonlinear index shift experienced by the probe, is the sum of the electronic and rotational responses [19],

$$\Delta n_p(x,t) = 2n_2 I(x,t) + \int_{-\infty}^{t} R(t-t')I(x,t')dt' \quad (2)$$

where $I$ is the pump intensity, $n_2$ is the electronic Kerr coefficient, which depends on the probe and pump wavelengths $\lambda_p$ and $\lambda_e$, and $R$ is the impulse response function for quantized rotations of a rigid rotor. The terms in Eq. (2) are related to our measured phase shifts by

$$\Delta\varphi_{elec}(x,t) = 2n_2 I_0 f(x,t) k_{pr} L$$

$$\Delta\varphi_{rot}(x,t) = \frac{2\pi N k_p L}{n_0} \Delta\alpha(\lambda_p)(<\cos^2\theta>_t - \tfrac{1}{3})$$

$$= \frac{2\pi N k_p L}{n_0} \Delta\alpha(\lambda_p) \frac{\Delta\alpha(\lambda_e) I_0}{c\hbar} \int_{-\infty}^{t} g(t-t') f(x,t') dt'$$

(3)

where $I_0$ is the pump peak intensity and $f(x,t)$ is its normalized spatiotemporal profile, $k_p$ is the probe wavenumber, $L \sim 2$ mm (< pump confocal parameter of ~5 cm) is the pump-probe interaction length, $N$ is the molecule number density, $n_0$ is the background gas refractive index, $\Delta\alpha(\lambda_e)$ ($\Delta\alpha(\lambda_p)$) is the polarizability anisotropy at the pump (probe) wavelength, $<\cos^2\theta>_t - \tfrac{1}{3}$ is the ensemble-averaged transient alignment induced by the pump pulse [24], and the rescaled impulse response function is [24]

$$g(t) = \frac{-16\pi}{15} \sum_j \frac{j(j-1)}{2j-1} (\rho_j^{(0)} - \rho_{j-2}^{(0)}) \sin\omega_{j,j-2} t \qquad (4)$$

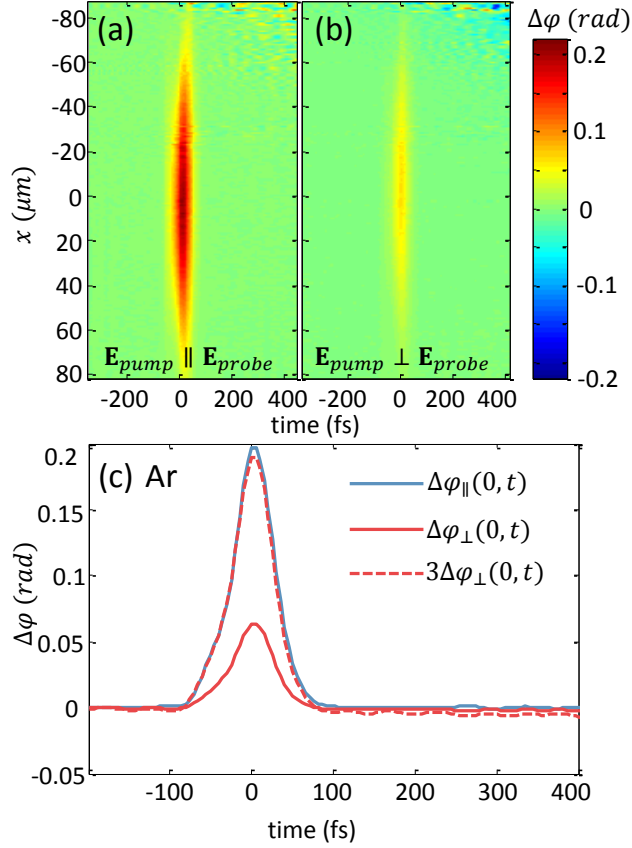

**Fig. 3.** Experimental results in Ar at 1250 nm pump central wavelength: (a) $\Delta\varphi_\parallel(x,t)$ and (b) $\Delta\varphi_\perp(x,t)$. (c) Temporal lineouts of the parallel and perpendicular polarized phase shift and the perpendicular phase shift scaled by a factor of 3.

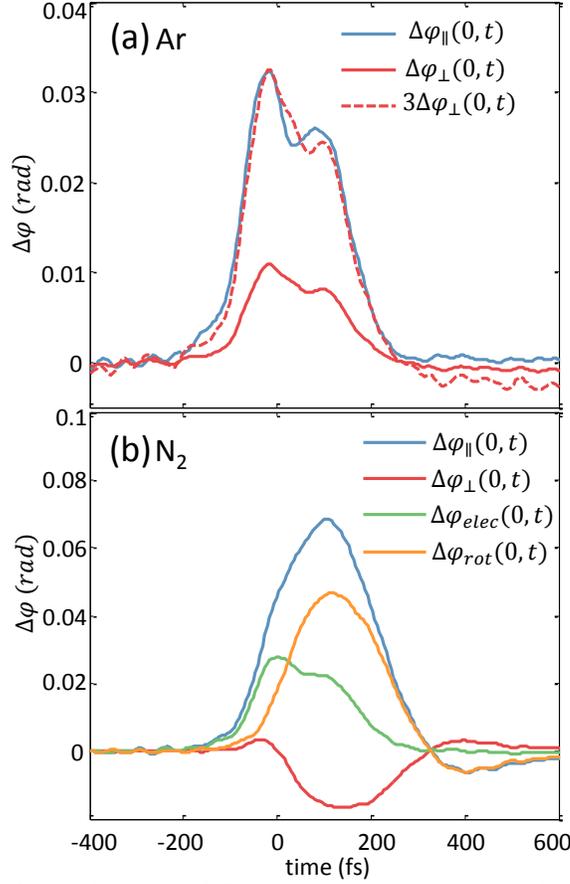

**Fig. 4.** Experimental results in (a) Ar and (b) $N_2$ at 2200 nm, a wavelength where the pump pulse shape is longer and more complex than at shorter wavelengths.

where $\rho_j^{(0)}$ is the thermal population of state $j$, $\omega_{j,j-2} = 4\pi B(2j-1)$, and $B$ is the rotational constant.

Essential to our method for absolute measurements is accurate recovery of the pump pulse envelope. Figures 4(a) and (b) show the nonlinear phase shift at $\lambda_e$ = 2200 nm for argon and nitrogen. The somewhat complex pulse pump envelope $f(x,t)$ measured in (a) results from the propagation of the OPA idler through a dichroic mirror. Figure 4(b) shows that such accurate recovery of the intensity envelope enables a clean separation of the electronic and rotational responses, even when the envelope is complex. Examination of Eq. (3) now shows that measurement of $\Delta\varphi_{rot}$ and computation of the convolution integral $\int_{-\infty}^{t} g(t-t')f(x,t')dt'$ from the known and measured functions $g$ and $f$ gives $\mu_1 = \Delta\alpha(\lambda_p)\Delta\alpha(\lambda_e)NLI_0$. This then allows determination of $\mu_2 \equiv n_2(\Delta\alpha(\lambda_p)\Delta\alpha(\lambda_e))^{-1}$ through the equation for $\Delta\varphi_{elec}(x,t)$. It is important to note that there is a large two-dimensional sample size of $\mu_2$ measurements in each SSSI shot, since the phase shift measurement is both time and 1D-space resolved. In the extracted phases shown here (for example, Fig. 2) there are ∼100 points in $x$ and ∼50-100 points in $t$, so in principle each shot embodies a maximum ∼5000 measurements of $\mu_2$.

Our results are summarized in Table 1. The uncertainties quoted in the table originate from three sources. First is the residual square error in $\mu_1$ from least squares fits of $\Delta\varphi_{rot}(x,t)$ (Eq. (3)) to the data points spanning $t$ for fixed $x$. The $x$-average of these results gives the tabulated $n_2$ values, and the standard deviation is one source of uncertainty. Another source of uncertainty is slight laser average power drift over the course of a run, and we include its estimated effect in the displayed error. Finally, we include the propagated uncertainty in the value of $\Delta\alpha$, as given in [19]. The noticeably higher error for the $\lambda_e$=2400 nm measurements originates from the very small nonlinear phase shifts (maximum of ~30 mrad) measured for that low peak pump intensity.

From the expression above for $\mu_2$ it is clear that determination of $n_2$ requires assessment of dispersion in $\Delta\alpha$. Such dispersion has been calculated to have the approximate frequency dependence $\Delta\alpha(\omega) \approx \Delta\alpha(0) + C\omega^2$ [28]. Light scattering measurements for $N_2$ [29] are well fit by $\Delta\alpha(0) = 6.6 \times 10^{-25}$ cm$^3$ and $C = 3.8 \times 10^{-57}$ cm$^3$s$^2$. This relation, with $\Delta\alpha(0)$ scaled so that $\Delta\alpha(\omega)$ matches our measured value of $\Delta\alpha$ at $\lambda_e$=800 nm [19], was used to account for the dispersion of $\Delta\alpha$ in the analysis. The dispersion is quite weak: $\Delta\alpha(\omega_{800nm})/\Delta\alpha(0)$ ~1.03 and $\Delta\alpha(\omega_{400nm})/\Delta\alpha(0)$ ~1.13.

It is evident from Table 1, that neglecting $\Delta\alpha$ dispersion affects the results by at most ~10% near $\lambda$=400 nm and <2% at longer wavelengths (compare columns (a) and (b)). Next, we see that results for $n_2$ are dispersionless within the precision of our measurements, except near $\lambda_e$=400 nm (except for $O_2$). This is in accord with calculations of the third order hyperpolarizability at optical frequencies below electronic resonances [21, 30]. It is important to be clear that while the Kerr coefficient we actually desire is $n_2(\omega_e) = (12\pi^2/n_0^2 c)N\gamma^{(3)}(\omega_e;\omega_e,-\omega_e,\omega_e)$, where $\gamma^{(3)}$ is the third order hyperpolarizability, our pump-probe experiment actually measures $n_2(\omega_e,\omega_p) \propto \gamma^{(3)}(\omega_p;\omega_e,-\omega_e,\omega_p)$. However, the $\gamma^{(3)}$ dispersion formula of [21] and its parametrization by Shelton and Rice [30] give

$$\gamma^{(3)}(\omega_s;\omega_1,\omega_2,\omega_3) = \gamma_0\left(1 + a\omega_L^2 + b\omega_L^4 + ...\right), \quad (5)$$

where $\omega_s = \omega_1 + \omega_2 + \omega_3$, $\gamma_0$ is the static hyperpolarizability, $\omega_L^2 = \omega_s^2 + \omega_1^2 + \omega_2^2 + \omega_3^2$, and $a$=1.8×10$^{-33}$, 3×10$^{-33}$ s$^2$, $b$=1.5×10$^{-65}$, 1.6×10$^{-65}$ s$^4$ and $\gamma_0$= 7.7×10$^{-30}$, 9.8×10$^{-30}$ cm$^6$W$^{-1}$s$^{-1}$ for $N_2$ and Ar respectively. For our pump-probe case, $\omega_L^2 = 2\omega_e^2 + 2\omega_p^2$ and Eq. (5) gives, at worst (shortest $\lambda_p$ and longest $\lambda_e$), $\delta = \left|(n_2(\omega_e,\omega_p) - n_2(\omega_e))/n_2(\omega_e)\right| < ~0.06$ for pump and probe wavelengths longer than ~500 nm, which applies to most of Table I and is within our measurement error. For $\lambda_e$= 400 nm and $\lambda_p$=600 nm (peak of SC), $\delta <$ ~0.06, also within our measurement error, confirming the dispersion of $n_2$ in $N_2$ and Ar near 400 nm.

In Table 1, we compare our results in Ar, $N_2$, and $O_2$ with the values of $n_2$ calculated with Eq. (5) using values of $\gamma_0$, $a$, and $b$ (shown above for $N_2$ and Ar) found using the electric field induced second harmonic generation (ESHG) technique [31], in which $\omega_1 = \omega_2 = \omega$ and $\omega_3 = 0$. The agreement is very good. The ESHG results in argon and nitrogen [31] have been shown to agree with theoretical calculations [32,33] to within ~10%. Both theory and ESHG experiments agree with our finding that the nonlinear refractive index is quite dispersionless in the infrared. For application of these results to propagation simulations, both the electronic and rotational contributions must be considered as shown in Eq. (2). For short pump pulses in $N_2$ and $O_2$ (<~50 fs), the electronic response dominates and for longer pulses (>~150 fs), the rotational response dominates, as inferred directly from Figs. 2 and 4. In the limit of a very long pulse, the molecular response is adiabatic, and the effective nonlinearity coefficient can be written as $n_{2,eff} = n_2 + n_{2,rot}$, where calculated values of $n_{2,rot}$ (using our measured values of $\Delta\alpha$ [19]) are shown in Table 1.

In conclusion, we have used single shot supercontinuum spectral interferometry to measure the electronic Kerr coefficients for the major atmospheric constituents, $N_2$, $O_2$ and Ar, at wavelengths ranging from 400 nm to 2400 nm. Our measurements are referenced to the polarizability anisotropy of the molecular gases, which enables extraction of absolute nonlinearities without the need for separate measurements of the gas density or pump intensity profiles. Except for the nitrogen and argon measurements at pump wavelength near 400 nm, the Kerr coefficients are measured to be dispersionless within the precision of our apparatus and consistent with the theoretical predictions.

|  | (a) N$_2$ | | | (b) N$_2$ | | | (c) O$_2$ | | | (d) Ar | |
|---|---|---|---|---|---|---|---|---|---|---|---|
| Pump wavelength $\lambda_e$ (nm) | $n_2$ ($10^{-20} cm^2/W$) | | | $n_2$ ($10^{-20} cm^2/W$) | | | $n_2$ ($10^{-20} cm^2/W$) | | | $n_2$ ($10^{-20} cm^2/W$) | |
|  | This work | [28] | $n_{2,rot}$ | This work | [28] | $n_{2,rot}$ | This work | [28] | $n_{2,rot}$ | This work | [28] |
| 400 nm | 10.1±1.2 | 9.4 | 29 | 9.3±1.1 |  | 24 | 8.5±0.8 | 10.8 | 54 | 10.9±1.3 | 12.2 |
| 800 nm | 7.9±0.8 | 8.4 | 24 | 7.9±0.8 |  | 24 | 8.1±0.7 | 9.0 | 54 | 10.1±1.0 | 10.7 |
| 1250 nm | 7.7±0.7 | 8.2 | 23 | 7.9±0.8 |  | 24 | 8.9±0.7 | 8.8 | 54 | 10.5±1.4 | 10.5 |
| 1650 nm | 8.0±0.8 | 8.1 | 23 | 8.1±0.8 |  | 24 | 7.9±0.6 | 8.7 | 54 | 10.9±1.0 | 10.4 |
| 2200 nm | 7.2±0.8 | 8.1 | 23 | 7.4±0.8 |  | 24 | 8.2±0.8 | 8.6 | 54 | 9.3±1.0 | 10.4 |
| 2400 nm | 7.6±1.3 | 8.1 | 23 | 7.8±1.3 |  | 24 | 10.0±1.2 | 8.6 | 54 | 9.9±1.7 | 10.3 |

(a) $n_2$ measurements are adjusted for $\Delta\alpha$ dispersion using coefficients found in [29]

(b) $n_2$ measurements are not adjusted for $\Delta\alpha$ dispersion. We used $\Delta\alpha(\omega_{800nm})$, measured in [19], for all wavelengths.

(c) $n_2$ measurements are not adjusted for $\Delta\alpha$ dispersion.. We used $\Delta\alpha(\omega_{800nm})$, measured in [19], for all wavelengths.

(d) Referenced to N$_2$ measurements.

Table 1. Measured values of $n_2$ ("This work") of major air constituents, scaled to atmospheric pressure. Comparison is shown to values of $n_2$ calculated for our wavelengths using Eq. (5), where the coefficients $a$, $b$, and $\gamma_0$ were measured using the ESHG technique [28]. For long pulses, the effective nonlinearity coefficient can be written as $n_{2,eff} = n_2 + n_{2,rot}$.

The authors thank Eric Rosenthal, Nihal Jhajj, and Ilia Larkin for useful discussions and technical assistance. The authors also thank M. Kolesik for theoretical discussions. This work was supported by the Air Force Office of Scientific Research (AFOSR), the Army Research Office (ARO), and the National Science Foundation (NSF).